\documentclass[aps, prd, reprint,showpacs,nofootinbib]{revtex4-1}

\usepackage{amssymb, amsmath, braket, nicefrac, cancel, graphicx} 
\usepackage[tight]{subfigure}
\usepackage{relsize}

\newcommand{\I}{\ensuremath{\mathbb{I}}}
\newcommand{\V}{\ensuremath{\mathbb{V}}}
\newcommand{\Amplitude}{\ensuremath{\mathcal{A}(\nu_\mu \rightarrow \nu_e)}}

\newcommand{\numu}{\ensuremath{\nu_\mu } }                   
\newcommand{\nue}{\ensuremath{\nu_e } }                      
\newcommand{\anumu}{\ensuremath{\overline{\nu}_{\mu}} }      

\newcommand{\anue}{\mbox{$\overline{\nu}_{e}$}}         

\begin{document}

\title{\bf  Constraints for nonstandard interaction $\epsilon_{e \tau}V_e$  from
$ \nue$ appearance in MINOS and T2K}

\author{Jo\~{a}o A. B. Coelho, Tomas Kafka, W. Anthony Mann, Jacob Schneps, and Ozgur Altinok}

\affiliation{Tufts University, Medford, MA 02155}


\begin{abstract}
Event rates for $\nue$ and $\anue$ appearance oscillations reported by the MINOS and T2K long baseline experiments
are used to set constraints on the strength of the non-standard interaction $\epsilon_{e \tau}V_e$ matter potential. 
The ranges allowed for the magnitude and phase of $\epsilon_{e \tau}$ are delineated for scenarios wherein {\it i)} other
non-standard interactions for neutrinos propagating through the terrestrial crust are negligible, and  {\it ii)} the real-valued,
flavor-diagonal couplings $\epsilon_{ee}$ and $\epsilon_{\tau \tau}$ are also operative.   Our analysis makes use of accurate
analytic forms for the $\nue$ amplitude $\mathcal{A}(\nu_\mu \rightarrow \nu_e)$ describing neutrino oscillation 
in constant-density matter in the presence of $\epsilon_{e \tau}$,  $\epsilon_{ee}$ and $\epsilon_{\tau \tau}$ non-standard interactions.
\end{abstract}

\maketitle


\section{Introduction}

Within the span of one year, multiple independent neutrino oscillation measurements by
reactor and accelerator long-baseline experiments have established the neutrino
$\theta_{13}$ mixing angle to be $\sim 9^{\circ}$~\cite{ref:T2K-big-theta13, ref:MINOS-improved-nue,
ref:Double-Chooz, ref:Daya-Bay, ref:RENO}.    For neutrino oscillation experiments, the
newly gained knowledge of the $\sin^2 2\theta_{13}$ mixing strength brings significant clarifications concerning sensitivity 
to CP-violating effects as may ensue with either the normal hierarchy (NH) or the inverted hierarchy (IH) 
for the neutrino mass eigenstates.

The revelation of $\theta_{13}$ has enabled first-time determinations of
``exclusion curves'' for the Dirac CP phase, $\delta$,  for each mass hierarchy in
experiments having sensitivity to perturbations in neutrino oscillations arising
from terrestrial matter effects:  The SuperKamiokande collaboration has reported $\chi^{2}$ versus 
$\delta$ fits for NH and for IH using atmospheric neutrino data~\cite{ref:Itow-Kyoto};   exclusion confidence level (C.L.) curves are reported
 by the MINOS collaboration based upon $\nue$ and $\anue$ appearance at the experiment's 735-kilometer baseline~\cite{ref:Vahle-FNAL}.
While the exclusion levels thus far achieved are quite modest, they serve to remind that a new era of experimental scrutiny of
neutrino flavor oscillations is getting underway and it remains to be seen whether conventional three-flavor mixing phenomenology
will continue to be an adequate framework:  
observational deviations from this framework are an exciting possibility as plausible harbingers of physics beyond the Standard Model.
For example,  the possibility that neutrinos propagating through dense matter 
may participate in effective, neutral-current like nonstandard interactions (NSI) has received considerable attention for more than a decade~\cite{ref:Ohlsson-pp2012}.

In standard three-flavor neutrino oscillation phenomenology, the Hamiltonian in flavor basis includes the
Mikheyev-Smirnov-Wolfenstein (MSW) matter potential.   The MSW potential accounts for coherent forward scattering of electron-flavor neutrinos 
from the electrons of ambient matter~\cite{ref:MSW}.    In an NSI scenario, the Hamiltonian carries additional matter potential terms analogous
to the MSW potential which allow flavor-changing as well as flavor-conserving NSI scattering processes.
There are six possible NSI amplitudes which can arise in neutrino propagation through matter.    These include three real-valued, flavor-diagonal
amplitudes conventionally designated as  $\epsilon_{ee}$, $\epsilon_{\mu\mu}$, and $\epsilon_{\tau\tau}$, and three flavor-changing amplitudes
which may carry CP-violating phases: $\epsilon_{e\mu}$, $\epsilon_{e\tau}$, and $\epsilon_{\mu\tau}$.
The phenomenology of neutrino NSI in propagation for accelerator, atmospheric, and solar neutrinos and for
various beam-plus-detector(s) configurations, has received extensive treatment.   Data from hadron and lepton colliders have also been utilized
in NSI studies, for NSI couplings involving quarks or electrons can give rise to anomalous monojet, monophoton, and multilepton events~\cite{ref:ATLAS-NSI}.
Comprehensive citations to the published
literature can be found in~\cite{ref:Ohlsson-pp2012, ref:Minakata, ref:Yasuda, ref:Parke-2010}.

The available data allow upper bounds to be set on
the magnitudes of NSI couplings.  According to the analysis of Ref.~\cite{ref:Biggio}, the effective NSI parameters for terrestrial matter are 
$|\epsilon_{\mu \mu}| < 0.07$, $|\epsilon_{e \mu}|< 0.33$, 
and $|\epsilon_{\mu \tau}| < 0.33$ at 90$\%$ C.L.
Additionally, on the basis of consistency with the high-energy atmospheric data, it is proposed that
bounds of a few percent are appropriate for $|\epsilon_{ e \mu}|$ and $|\epsilon_{\mu \tau}|$~\cite{ref:Yasuda, ref:Maltoni-2011}.   
For $|\epsilon_{e e}|$,  $|\epsilon_{e \tau}|$, and 
$|\epsilon_{\tau \tau}|$ however the bounds at 90$\%$ C.L. 
are much weaker; Ref.~\cite{ref:Biggio} finds these to be $<$ 4.2, $<$ 3.0, and $<$ 21 respectively.  
For neutrino NSI with electrons (but not with u or d quarks), more stringent limits 
have been set for  $\epsilon^{eR}_{ee}, \epsilon^{eL}_{ee}, \epsilon^{eR}_{\tau \tau}$, and $\epsilon^{eL}_{\tau \tau}$
based upon analysis of solar and KamLAND neutrino data~\cite{ref:Bolanos}.
With respect to NSI limits derived using monojet plus missing energy data sets of hadron collider experiments, our characterization
is appropriate for the ``light mediator" regime of anomalous monojet processes involving NSIs ~\cite{ref:ATLAS-NSI, ref:Friedland-pp2012}.
Thus, at present, the $\epsilon_{e\tau}$, $\epsilon_{ee}$, and $\epsilon_{\tau\tau}$ NSI are so poorly constrained that
large matter effects, of strengths which rival or exceed that of the MSW matter potential, remain as viable
phenomenological possibilities.
These three couplings enter into the probabilities for $\nue$ appearance oscillations, and the perturbations
they may introduce can be searched for by observing $\numu$ and $\anumu$ beams at long baselines.

Concerning $\epsilon_{e \tau} V_e$, the current situation with the solar $^{8}$B neutrino energy spectrum is worthy of note~\cite{ref:Friedland-2004, ref:Palazzo-Valle}.
The low-threshold measurements carried out thus far by Borexino, Super-Kamiokande, and the Sudbury Neutrino Observatory 
do not exhibit an upturn with decreasing $E_{\nu}$ as predicted by standard oscillations with the MSW matter effect.    However a deviation
from conventional oscillations in the form of a flatter spectrum is naturally provided 
by an $\epsilon_{e \tau}$ coupling strength $\mathcal{O}(10^{-1})$~\cite{ref:Palazzo-RapidC, ref:Friedland-pp2012}.

In this work we examine manifestations of the NSI matter potential 
$\epsilon_{e \tau} V_e$ together with $\epsilon_{e e} V_e$ and $\epsilon_{\tau \tau} V_e$
as may occur in neutrino propagation through the
constant-density terrestrial crust.    We focus on 
$\nue$ and $\anue$ appearance oscillations and evaluate the implications of recent
signal event counts reported by the T2K (295 km)~\cite{ref:T2K-big-theta13, ref:Nakaya-T2K} and MINOS (735 km)~\cite{ref:MINOS-improved-nue, ref:Vahle-FNAL, ref:Nichol-Kyoto} long-baseline experiments.
The sensitivity of conventional, terrestrial long-baseline experiments to the $\epsilon_{e \tau} V_e$ NSI has been explored in previous works
by other researchers.    In particular, experimental $\epsilon_{e \tau} V_e$ sensitivity has been examined for the 295-kilometer baseline of T2K~\cite{ref:Adhikari},
for the 735-kilometer baselines of MINOS and OPERA~\cite{ref:Friedland, ref:Yasuda-2007, ref:Sugiyama-2008, ref:BOS, ref:Huber-2008}, for the 810-kilometer baseline 
of NO$\nu$A~\cite{ref:Friedland-pp2012}, and for the
1050 km baseline proposed for T2KK~\cite{ref:Yasuda}.   Several of these studies make use of constraints
deduced from testing NSI scenarios using data from atmospheric neutrino experiments~\cite{ref:FLM04, ref:FL05, ref:Maltoni-2011}.
The work reported here utilizes the insights from these previous studies and examines the most recent observations of
positive $\nue$ appearance in two accelerator beam long-baseline experiments in light of the newly delineated value range allowed to $\theta_{13}$.
Our treatment is restricted to neutral current NSI processes as may occur with neutrino propagation in matter.   Complications arising
from possible NSI effects in neutrino production and/or detection processes in current experiments are not considered~\cite{ref:Ohlsson-pp2012}.

\section{Outline} \vspace{-9pt}

We proceed as follows:   In Sec. III we define the matter Hamiltonian to include
the NSI $\epsilon_{e\tau}$, $\epsilon_{ee}$, and $\epsilon_{\tau\tau}$, and we
assume, on the basis of bounds previously proposed~\cite{ref:Biggio, ref:Yasuda} that the
$\epsilon_{\mu\mu}$, $\epsilon_{e\mu}$, and $\epsilon_{\mu\tau}$ are much smaller and 
can be neglected.  We then present a formalism which characterizes three-flavor neutrino
oscillations with NSI.   Specifically, we express the $\nue$ appearance
amplitude $\mathcal{A}(\nu_\mu \rightarrow \nu_e)$ as a sum of three terms, $T_{i}$ ($i$=1,2,3), the 
absolute square of which gives the appearance oscillation probability for neutrinos
traversing terrestrial matter of constant density.    Our expressions are obtained by deriving the
time evolution operator for which the three-flavor Hamiltonian including matter effects is the
generator.    The methodology for this approach is presented in Ref.~\cite{ref:Exact-Form};
a summary of the derivation with inclusion of the NSI considered here is given in the Appendix below.
The analytic forms serve to illuminate the relative contributions arising from
the various NSI and from the CP phases $\delta$ and $\delta_{e \tau}$.  
Their compact nature is effective in reducing
input for computation, thereby increasing algorithm speeds.
The analytic forms have been used to check our fits to the data which are carried out using
numerical techniques.

In Sec. IV we summarize the observations of the MINOS and T2K experiments that we
use in order to fit for the NSI couplings.   In Sec. V we present a sequence of NSI fits to the data.
We commence with a minimalist scenario, namely that $\epsilon_{e \tau}$
is the only active NSI and that its coupling is real-valued (hence neglecting its phase $\delta_{e \tau}$
degree of freedom).    We then fit for the magnitude $|\epsilon_{e \tau}|$ and the sum of the 
CP phases  $\delta + \delta_{e \tau}$.     Finally we consider the realistic situation wherein
complex $\epsilon_{e \tau}$ is active together with the flavor-diagonal NSI couplings
$\epsilon_{ee}$, and $\epsilon_{\tau\tau}$.     For the latter we introduce a constraining
relationship which is based upon the behavior of $\numu$ disappearance oscillations
at high energies for atmospheric neutrinos.    This allows the number of variables in 
the fit to be limited to $|\epsilon_{e \tau}|$, $\epsilon_{ee}$, together with the above-mentioned
sum of CP phases.    In Sec. VI we summarize the constraints for $|\epsilon_{e \tau}|$  which
are indicated by our fits, and take note of near-term experimental developments which
will enable these constraints to be improved.

\section{Amplitude for $\numu \rightarrow \nue$ Oscillations}
\label{sec:Exact-Amplitude}
\subsection{Three-flavor oscillations with NSI matter effects}
\label{subsec:Exact-intro}

For neutrino propagation in vacuum, the Hamiltonian in the basis of three mass eigenstates
$\nu_i$ ($i$ =1,~2,~3) is 
\begin{equation}
\hat H_0^{(i)} 
 = \frac{1}{2\ell_{v}} \cdot \mathrm{diag}   \left( 0, \alpha , 1\right) , ~~\text{where}
 \end{equation} 
\begin{equation} 
 \ell_v \equiv \frac{E_\nu}{\Delta m_{31}^2} ~~~\text{and} ~~\alpha \equiv \frac{\Delta m^2_{21}}{\Delta m^2_{31}}
 \end{equation} 
 are, respectively,  the vacuum oscillation length and the mass hierarchy ratio.
The transformation from mass basis $\{\ket{\nu_i}\}$ to neutrino flavor basis $\{\ket{\nu_{\varphi}}\}$ 
($\varphi=e,~\mu,~\tau$) is provided by the unitary mixing matrix
\begin{equation} \label{eq:mixing_matrix}
  \hat U_{mix} \equiv \hat R_1(\theta_{23}) \cdot \mathbb{\hat I}_{\delta} \cdot \hat R_2(\theta_{13}) \cdot 
  \mathbb{\hat I}_{-\delta} \cdot \hat R_3(\theta_{12})
\end{equation}
wherein the atmospheric and solar mixings are accounted for via the rotation matrices $\hat R_1(\theta_{23})$ and $\hat R_3(\theta_{12})$.
The Dirac CP phase $\delta$ is included via the auxiliary matrices $\mathbb{\hat I}_{\delta} \equiv \mathrm{diag}(1, 1, e^{i\delta})$ and 
$ \mathbb{\hat I}_{-\delta}  =  \mathbb{\hat I}_{\delta}^\dagger $.
Then the vacuum Hamiltonian in flavor basis is given by the unitary transformation
\begin{equation}
\hat  H_0^{(\varphi)}
    = \hat U_{mix} \hat H_0^{(i)} \hat U_{mix}^\dag ~,
\end{equation}
and the effective wave equation for vacuum propagation of flavor states is
\begin{equation} \label{eq:schroedinger_vacuum}
  i \frac{d}{dt} \vec \nu^{(\varphi)}(t) = \hat H_0^{(\varphi)} \vec \nu^{(\varphi)}(t).
\end{equation}
To $\hat H_0^{(\varphi)}$ we add (in flavor basis) the MSW and NSI matter interactions:
\begin{equation}\label{eq:matter-interactions-1}
 \hat H_{\mathrm{matter}}^{(\varphi)}
    = V_{e} \left(
      \begin{array}{ccc}
        1+ \epsilon_{ee} & 0 & \epsilon_{e \tau}  \\
        0 & 0 & 0 \\
         \epsilon_{e \tau}^{*} & 0 &  \epsilon_{\tau \tau}
    \end{array} \right).
\end{equation}
Here, $V_{e} = \sqrt{2} G_{F} n_{e}$ is the MSW matter interaction \cite{ref:MSW}  where $G_{F}$ is the Fermi coupling constant and $n_{e}$ is
the electron density in matter.   The standard MSW matter effect is modified by the presence of the real-valued, diagonal NSI
interactions $\epsilon_{ee} V_{e}$ and $\epsilon_{\tau \tau} V_{e}$, and by the off-diagonal $\epsilon_{e \tau} V_{e}$ interaction.   Since the
latter amplitude may carry a CP-violating phase, $\delta_{e \tau}$, hereafter we designate the magnitude $|\epsilon_{e \tau}|$
and display the phase explicitly.    It is convenient to absorb $V_{e}$ into the matter potential,  
\begin{equation}
 A \equiv 2 \ell_{v} V_{e} , 
\end{equation}
and to write Eq.~\eqref{eq:matter-interactions-1} as
\begin{equation}\label{eq:matter-interactions-2}
\hat  H_{\mathrm{matter}}^{(\varphi)}
     = \frac{A}{2\ell_v} \left(
      \begin{array}{ccc}
        1+ \epsilon_{ee} & 0 & |\epsilon_{e \tau}|  e^{i\delta_{e \tau}} \\
        0 & 0 & 0 \\
         |\epsilon_{e \tau}| e^{-i\delta_{e \tau}} & 0 &  \epsilon_{\tau \tau}
    \end{array} \right).
\end{equation}

A method to solve the time evolution operator in flavor basis $ \hat U^{(\varphi)}(t = \ell,0)$ for propagation
to baseline distance $\ell$ in constant density matter is presented in Ref.~\cite{ref:Exact-Form}, for conventional three-flavor oscillations with 
$ \hat H_{\mathrm{matter}}^{(\varphi)} = \mathrm{diag} \left( A/2\ell_{v}, 0, 0 \right) $.   This same approach can be used
to obtain an accurate solution for the more elaborate matter interactions of Eq.~\eqref{eq:matter-interactions-2}.    Following 
Ref.~\cite{ref:Exact-Form}, the matrix elements of the evolution operator  $ \hat U^{(\varphi)}( \ell)$ corresponding to 
$\hat  H_0^{(\varphi)} + \hat H_{\mathrm{matter}}^{(\varphi)}$
provide the various possible three-flavor oscillation amplitudes.   The $\nu_{e}$ appearance amplitude is
given by element $\hat U^{(\varphi)}_{12}$ which can be broken out as a sum of three terms:
\begin{equation}\label{eq:Amplitude-Sum}
 \mathcal{A}(\nu_\mu \rightarrow \nu_e) =  T_{1} + T_{2} + T_{3} .
 \end{equation}
The component amplitudes $T_{i}$ comprise an analytic foundation for our investigation of NSI constraints
arising from the recent $\nu_{e}$ appearance observations by MINOS and T2K.

In the Sections to follow we specify the $T_{i}$ and then focus upon their implications.   Details concerning the derivation
of the evolution operator $ \hat U^{(\varphi)}( \ell)$ which underwrites the $T_{i}$ are provided in the Appendix.

\subsection{Specification of the $T_{i}$ amplitude terms}
\label{sec:Specify-Ts}

In order to write compact expressions for the $T_{i}$, we define some notations.   For the mixing angles we use $s_{ij} \equiv \sin\theta_{ij}$
and $c_{ij} \equiv \cos\theta_{ij}$; for the atmospheric oscillation phase we write
\begin{equation}\label{eq:Define_Delta}
    \Delta \equiv  \frac{\Delta m_{31}^2~ \ell}{4 E_\nu} =  \frac{\ell}{4\ell_v} ~.
\end{equation}
We define scaled forms $\alpha'$ and $\alpha''$ for the hierarchy parameter :
\begin{eqnarray}
\label{eq:alpha}
 \alpha' ~&\equiv& \sin 2\theta_{12} \cdot \alpha~, ~~\text{and} \\
 \alpha'' &\equiv& (1-3 c_{12}^2) \cdot \alpha~. \nonumber
\end{eqnarray}
The mixing strengths involving $\theta_{13}$ are often accompanied by the factor $(1 - s_{12}^2\alpha)$, hence we define
\begin{eqnarray}
 \sin 2 \tilde\theta_{13} &=&(1 - s_{12}^2\alpha) \cdot  \sin 2 \theta_{13} , ~~\text{and} \nonumber \\
 \cos 2 \tilde\theta_{13} &=& (1 - s_{12}^2\alpha) \cdot \cos 2 \theta_{13} .
\end{eqnarray}
We need to refer to the elements of the full Hamiltonian in propagation basis as obtained after a re-phasing
of its diagonal elements (see Sec. IV.B of Ref.~\cite{ref:Exact-Form}).
The Hamiltonian at that stage has the form
\begin{eqnarray}\label{eq:elements-Hp}
  \hat H^{(p)}
    &=& \left(
      \begin{array}{ccc}
        -Q & r & f \\
        r^* & -G & b \\
        f^* & b^* & +Q
      \end{array} \right)
 \end{eqnarray}
Its diagonal elements are the following real-valued functions:
\begin{equation}\label{eq:Define-Q-G}
\begin{split}
  Q &\equiv \frac{1}{4\ell_v} \left(\cos 2\tilde\theta_{13} - A \left[ 1 + \epsilon_{ee}  - c_{23}^{2} \epsilon_{\tau \tau} \right] \right), \\
  G &\equiv \frac{1}{4\ell_v} \left(1+A \left[1 + \epsilon_{ee} -(2 s_{23}^2 - c_{23}^{2} ) \epsilon_{\tau \tau}  \right] + \alpha'' \right), \\
\end{split}
\end{equation}

We designate the complex-valued off-diagonal elements using lower-case letters as follows:
\begin{eqnarray}\label{eq:Define-frb}
  f &\equiv& \frac{1}{4\ell_v} \left( \sin 2\tilde\theta_{13} + 2 c_{23} |\epsilon_{e \tau}| e^{i(\delta + \delta_{e \tau})} A \right), \\
  r &\equiv& \frac{1}{4\ell_v} \left( c_{13} \alpha' - 2 s_{23} |\epsilon_{e \tau}| e^{i \delta_{e \tau}} A \right), \\
  b &\equiv& \frac{1}{4\ell_v} \left( -s_{13} \alpha'  - s_{23} c_{23} \epsilon_{\tau \tau} e^{i \delta} A      \right). 
\end{eqnarray}
Then we have
\begin{equation} \label{eq:Define-T1}
  T_1 = (-i) s_{23} \frac{f}{N} \cdot \sin(\bar{N} \Delta) \cdot e^{-i\delta}
\end{equation}
where
\begin{equation}\label{eq:define-N-bar}
\bar{N} \equiv 4\ell_{v} \cdot N \equiv 4\ell_{v} \cdot \left[ |f|^{2} + Q^{2} \right]^{\frac{1}{2}} .
\end{equation}
The second term in Eq.~\eqref{eq:Amplitude-Sum} is
\begin{equation}\label{eq:Define-T2}
  T_2 = (-i) c_{23} \frac{r}{\eta} \cdot \sin (\bar \eta \Delta ) \cdot e^{i\bar G \Delta} ,
\end{equation}
where 
\begin{equation}\label{define-eta-bar}
\bar{\eta} \equiv 4\ell_{v} \cdot \eta \equiv 4\ell_{v} \cdot \left[ |r|^{2} + |b|^{2} \right]^{\frac{1}{2}} ,
\end{equation}
and
\begin{equation}\label{define-G-bar}
\bar G  \equiv 4\ell_{v} \cdot G .
\end{equation}

Of the three amplitude terms in Eq.~\eqref{eq:Amplitude-Sum}, $T_3$ is the most intricate.   If the NSIs were known to be
small, e.g. $|\epsilon_{\varphi \varphi'}| \leq \alpha$, then $T_3$ could be neglected.   However
large NSIs are a distinct possibility and so $T_3$ is to be
retained.    For convenience we define two complex functions $S_{1}$ and $S_{2}$:
\begin{equation} \label{eq:S1-def}
  S_1 \equiv \frac{rb}{\eta^2},
    \end{equation}
and
\begin{equation} \label{eq:S2-def}
 S_2 \equiv (-i) \left[ \frac{|r|^2}{\eta^2} \cdot \frac{f}{N} + S_1\cdot \frac{Q}{N} \right].
\end{equation}
$T_3$ can then be expressed as 
\begin{equation}\label{eq:Define-T3}
  T_3 = -2 s_{23} \cdot \left\{ S_1 \cos (\bar \eta \Delta ) + S_2 \sin (\bar \eta \Delta ) \right\}
        \cdot \sin^2( \frac{\bar \eta \Delta }{2}) \cdot e^{-i\delta}.
\end{equation}

\smallskip
In summary, the three amplitude terms of Eq.~\eqref{eq:Amplitude-Sum} are given by Eqs. \eqref{eq:Define-T1}, \eqref{eq:Define-T2}, and \eqref{eq:Define-T3}.
These can be coded as complex functions, and the $\nu_\mu \rightarrow \nu_e$ oscillation probability can be constructed as
 $|\mathcal{A}(\nu_\mu \rightarrow \nu_e)|^2$.

\subsection{Appearance probability upon neglecting $T_{3}$}

The ways in which NSI matter effects introduce distortions to conventional oscillations can be discerned in part
by examining an approximate form for the $\nue$ appearance probability $\mathcal{P}(\nu_\mu \rightarrow \nu_e)$.
Under the assumption that all $\epsilon_{\varphi \varphi'}$ are relatively small, we may neglect $T_3$ and write
\begin{equation}\label{eq:approx-P}
\begin{split}
 \left|\Amplitude \right|^2& \simeq \left|T_{1} + T_{2}\right|^{2} = s_{23}^2 \cdot |f|^2 \cdot \frac{\sin^2(\bar N \Delta )}{N^2} \\
  & + \sin 2 \theta_{23} \cdot \frac{\sin( \bar N \Delta)}{\bar N} \cdot
            \frac{\sin(\bar \eta \Delta)}{\bar \eta} \cdot \\
  & \Big\{
    c_{13}\sin 2\tilde\theta_{13} \cdot \alpha' \cdot \cos(\bar G \Delta+\delta) \\
  &
    - 2 s_{23}  \sin 2 \tilde\theta_{13} \cdot  |\epsilon_{e \tau}| A \cdot \cos(\bar G \Delta+\delta+\delta_m) \\
  &
    + 2 c_{23} c_{13} \cdot \alpha' \cdot |\epsilon_{e \tau}| A \cdot \cos(\bar G \Delta-\delta_m) \\
  &
    -2 \sin 2 \theta_{23} \cdot ( |\epsilon_{e \tau}| A)^2 \cdot \cos(\bar G \Delta)
  \Big\} \\
  & + c_{23}^2 \cdot  \frac{|r|^2}{\eta^2} \cdot \sin^2(\bar \eta \Delta).
\end{split}
\end{equation}
In the last term, the ratio $ \frac{|r|^2}{\eta^2}$ reduces to $c^2_{13}$ in the limit that the NSI couplings go to zero.
More generally, the oscillation probability of Eq.~\eqref{eq:approx-P} reduces to the three leading terms of the 
formula of Ref.~\cite{ref:Exact-Form} in the limit that the NSI interactions are turned off.    As discussed in Ref.~\cite{ref:Exact-Form},
these same three terms are related to the well-known perturbative formula of Cervera {\it et al.} (Ref.~\cite{ref:Cervera}; see
also \cite{ref:Freund}, \cite{ref:Akhmedov}).    One manifestation of a sizable $ |\epsilon_{e \tau}| $ occurs within the factor $|f|^2$ of the 
first term of Eq.~\eqref{eq:approx-P}.    Referring to Eq.~\eqref{eq:Define-frb}, one sees that the term containing 
$|\epsilon_{e \tau}| e^{i(\delta + \delta_{e \tau})} A$
~causes the effective mixing strength to deviate from $\sin 2 \tilde \theta_{13}$, giving a dependence upon $E_{\nu}$.
Amplitude expressions which lead to $\mathcal{P}(\nu_\mu \rightarrow \nu_e)$ of accuracy comparable to Eq.~\eqref{eq:approx-P} have been
discussed in previous works~\cite{ref:BOS, ref:Minakata}.

\section{$\nue$ Appearance in T2K and MINOS}

The occurrence of electron-shower dominated events with rates as predicted for $\numu \rightarrow \nue$ oscillations,
has recently been reaffirmed by the T2K  and MINOS long-baseline experiments.
In T2K, data exposures to the experiment's low-energy, off-axis ($2.5^{\circ}$) $\numu$ beam 
totaling $ 2.56 \times 10^{20}$ protons-on-target (PoT) have been analyzed. 
Among events having reconstructed energies less than 1250 MeV,  10 $\nue$ charged-current event candidates are observed, 
to be compared to $2.47$ background events predicted for null oscillations~\cite{ref:Nakaya-T2K}. 
The ten candidate signal events include six $\nue$ events reported previously by T2K as evidence for
a relatively large $\theta_{13}$ mixing angle~\cite{ref:T2K-big-theta13}.    

 The recent MINOS results are based on exposures to the NuMI low-energy beam of $ 10.6 \times 10^{20}$ PoT in neutrino-focusing mode
 and $ 3.3 \times 10^{20}$ PoT in antineutrino-focusing mode.   It is reported that, from data runs with $\numu$-focusing, 152 candidate $\nue$ events are observed while
 128.6 events are expected for null oscillations.  (For NH with $\sin^2 2\theta_{13} = 0.10$ and $\delta_{CP} = 0$, 161.1 events are expected.)
For running with $\anumu$-focusing (reversed horn-current running), 20 ($\anue + \nue$) candidate events are observed, while 17.5 events are expected for null oscillations.  (For NH with $\sin^2 2\theta_{13} = 0.10$ and $\delta_{CP} = 0$, 21.2 events are expected.)~\cite{ref:Nichol-Kyoto}.   
 
For the purpose of fitting to NSI scenarios, we treat both experiments as counting experiments
in which a signal has been measured over and above an estimated background.     Errors are assigned according to sample statistics plus allowance for
systematic errors associated with background estimation and signal detection.     For MINOS we allot a conservative systematic error estimate of $6\%$~\cite{ref:MINOS-improved-nue};  
for T2K we allot $15\%$~\cite{ref:T2K-big-theta13, ref:Nakaya-T2K}.   For the 295-km baseline of T2K and for the 735-km
baseline of MINOS as well,  neutrino propagation is confined to the Earth's crust, for which a density of $\rho = 2.72\; \mathrm{g/cm}^{3}$ is assumed.   In fitting of 
three-flavor neutrino oscillations with matter effects, we use $V_{e} = 1.1 \times 10^{-13}\;\mathrm{eV} = (1/1900)~\mathrm{km}^{-1}$~\cite{ref:OPERA}.

\section{Allowed Regions for $\epsilon_{e \tau}$}
 
 We proceed with fitting of three-flavor neutrino oscillations including NSI to the T2K and MINOS $\nue$ appearance data.
 A log-likelihood fit of three terms is used to compare the observed versus expected signal rates for $\nue$ appearance
in T2K, and for $\nue$ appearance and $\anue$ appearance in MINOS. 
 As previously noted, we neglect the NSI $\epsilon_{\mu\mu}$, $\epsilon_{e\mu}$, and $\epsilon_{\mu\tau}$   
on the basis of the existing upper bounds given in Sec. I, and focus on the possible role for $\epsilon_{e\tau}$ which may be operative 
in conjunction with $\epsilon_{ee}$  and $\epsilon_{\tau\tau}$.
Even with restriction to the latter three NSI, the number of degrees of freedom available to an oscillation 
scenario remains rather daunting, for the CP phases $\delta$ and $\delta_{e \tau}$ are present together with
the three coupling strengths, and the two possibilities for the mass hierarchy must be considered.
Additionally the fits require values to be specified for the atmospheric $\Delta m^2_{31}$and solar $\Delta m^2_{21}$ mass-squared differences 
and for the mixing angles $\theta_{23}$, $\theta_{13}$, and $\theta_{12}$.   For these
we use the world-average values and 1 $\sigma$ error ranges obtained for the normal hierarchy by Ref.~\cite{ref:Fogli-WA}.

In the fits, the probabilities are computed by constructing a numeric
Hamiltonian in flavor basis, solving for its eigensystem, and using it to
propagate the neutrino amplitudes.   
The oscillation probabilities are then assembled and multiplied by ``event densities" constructed so as to yield
the differential event rates predicted for null oscillations.   For each experiment, integration of the 
oscillation-weighted event density over the neutrino energy range probed provides the number of events
predicted in the presence of oscillations.   The prediction is then
compared to the observed number of events using the log-likelihood
distribution given below:   

\begin{equation}\label{chi-sq-1}
\chi^2 = - 2 \sum_{i=1}^{3} \ln \mathcal{L}(N^{i}_{p}\,|\, N^{i}_{obs}, \sigma_{p}^{i}) + \chi^2_{\rm penalty} ~,
\end{equation}
where

\begin{equation}\label{chi-sq-2}\nonumber
\begin{split}
&-\ln  \,\mathcal{L}(N_{p} \,|\, N_{obs},\sigma_p) = \\
&\min_{\xi}\left\{ N_{p}(1+\xi) - N_{obs}  + N_{obs}  \ln \left[\frac{N_{obs}}{N_{p}(1+\xi)} \right]                                   
+ \frac{(\xi N_{p})^2}{2\, \sigma_p^2} \right\}
\end{split}
\end{equation}
and
\begin{equation}\label{chi-sq-3}
\chi^2_{\rm penalty} = \frac{(s_{13} - {\bar s_{13}})^2}{{\delta \bar s_{13}}^2} + \frac{(s_{23} - {\bar s_{23}})^2}{{\delta \bar s_{23}}^2}.
\end{equation}
In the above expressions $N^{i}_{p}$ and $N^{i}_{obs}$ are the predicted and observed
number of events respectively, for experimental measurements $i=1,2,3$. 
The systematic
uncertainty of $N^{i}_{p}$ is denoted by $\sigma_{p}^{i}$ and is taken into
account by minimizing the nuisance parameter $\xi$
representing a fractional shift in $N^{i}_{p}$.  
 Current best-fit values for $\sin^2\theta_{13}$ and $\sin^2\theta_{23}$
are assigned to $\bar s_{13}$ and $\bar s_{23}$, and their uncertainties are given by
$\delta \bar s_{13}$ and $\delta \bar s_{23}$.    For all fits reported below, marginalization
is carried out for
$\sin^2 \theta_{23}$ and $\sin^2 \theta_{13}$~\cite{ref:Fogli-WA}.

\begin{figure}[!htb]
\includegraphics[width=8.5cm]{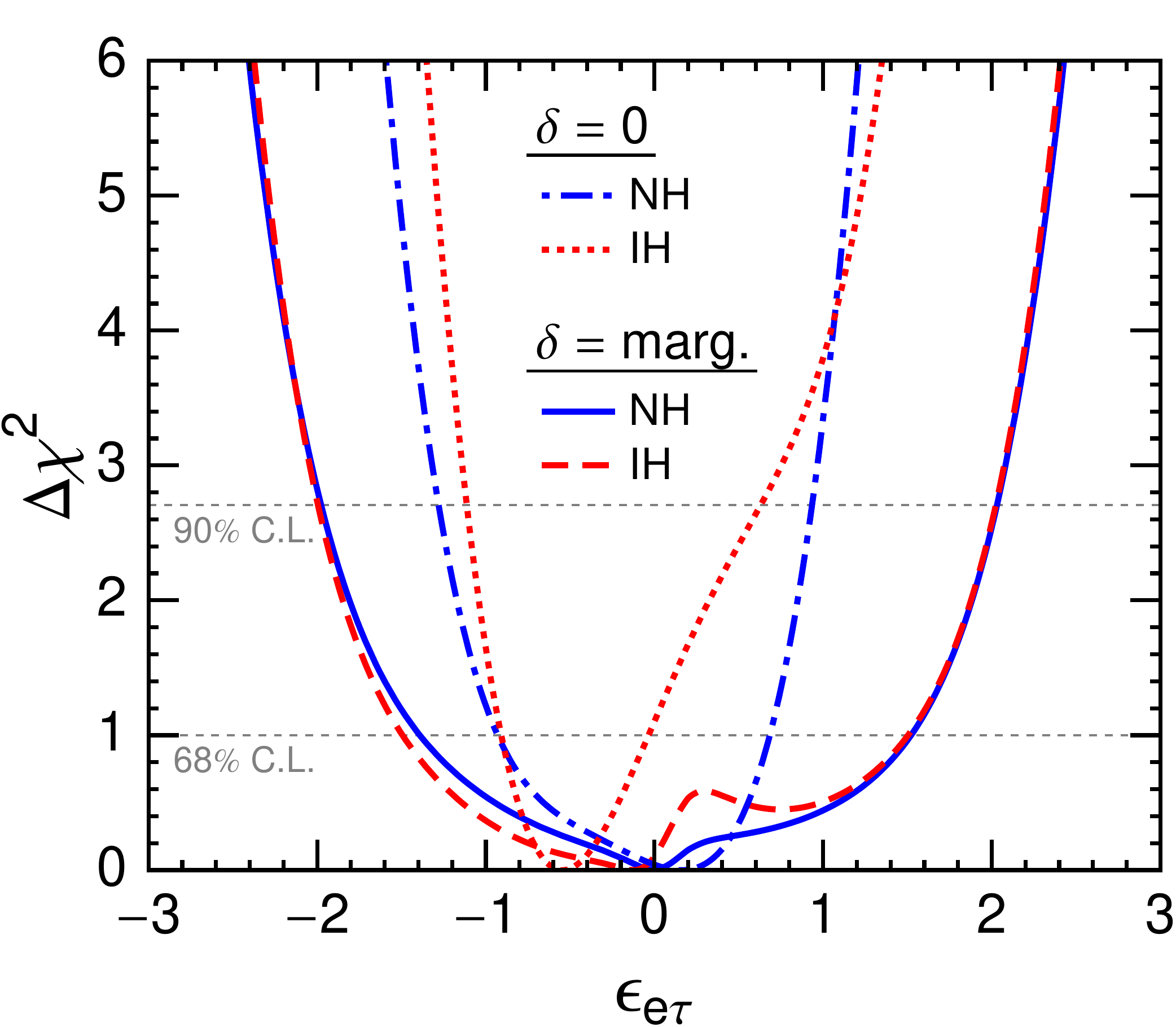}
\caption{\small Distributions of $\Delta \chi^2$ from fitting $\nue$ and $\anue$ appearance rates
reported by MINOS and T2K to $\numu (\anumu)  \rightarrow  \nue (\anue)$ oscillations
with the $\epsilon_{e \tau}$ NSI restricted to real values.   All fits are marginalized 
over the allowed ranges for the $\theta_{23}$ and $\theta_{13}$ mixing angles.   
For the fits of the dot-dash line (NH) and dotted line (IH) distributions, the Dirac CP phase, $\delta$, is
set to zero.   With $\delta$ marginalized in the fitting, the bounds on real-valued $\epsilon_{e\tau}$ become
less stringent as shown by the solid line (NH) and dashed line (IH) distributions.
}
\label{fig:Delta-chisq-vs-eps-etau}
\end{figure}

\subsection{Real-valued $\epsilon_{e \tau}$ as sole operative NSI}

As a first step, we consider a minimalist scenario in which $\epsilon_{e\tau}$ is the only operative NSI which we restrict 
to be real-valued, allowing it to be positive or negative but otherwise ignoring its phase degree of freedom.
We carry out two sets of fits with NH and IH treated separately in each set.   For the first set the value zero is assigned to the Dirac CP phase $\delta$, while 
for the second set the range 0 to $2\pi$ of $\delta$ is marginalized over in the fits.   In both sets the data is fitted to
$\numu (\anumu)  \rightarrow  \nue (\anue)$
oscillations with $\epsilon_{e\tau}V_{e}$ included together with the conventional MSW matter effect.
The distributions of $\Delta \chi^2 \equiv  \chi^2 -  \chi^2(\text{best-fit})$ from the two sets of fits are shown in
Figure~\ref{fig:Delta-chisq-vs-eps-etau}.    The distributions
serve as exclusion curves, with values of $\epsilon_{e\tau}$ having $\Delta \chi^2$ which exceed 1.0 (2.71) being
excluded at 68$\%$(90$\%$) C.L.   The first set of fits with the values of both CP phases assigned to zero, yield the dot-dash line ($\Delta m^2_{31} > 0$)
and dotted line ($\Delta m^2_{31} < 0$) distributions.   The distributions indicate $|\epsilon_{e\tau}| \leq 1.3$ at 90\% C.L. for either 
mass hierarchy.    For the same scenario but with solar scale mixing also neglected ($\alpha \rightarrow 0$), 
more stringent bounds have been obtained by fitting to atmospheric plus K2K neutrino oscillation data~\cite{ref:FLM04}.
However the bounds become distinctly more relaxed when the Dirac CP phase $\delta$ is accounted for via marginalization,
as shown by solid line (NH) and dashed line (IH) distributions from the second set of fits.    The latter exclusion curves are nearly identical
and so the hierarchies are not distinguished.    At $90\%$ C.L. our fits to real-valued $\epsilon_{e\tau}$ with $\delta$ marginalization yield the constraint
\begin{equation}\label{Real-e-tau-scenario}
   -2.0 < \epsilon_{e\tau} <  2.0~, 
\end{equation}
for either mass hierarchy.

\begin{figure}[!htb]
\includegraphics[width=8.5cm]{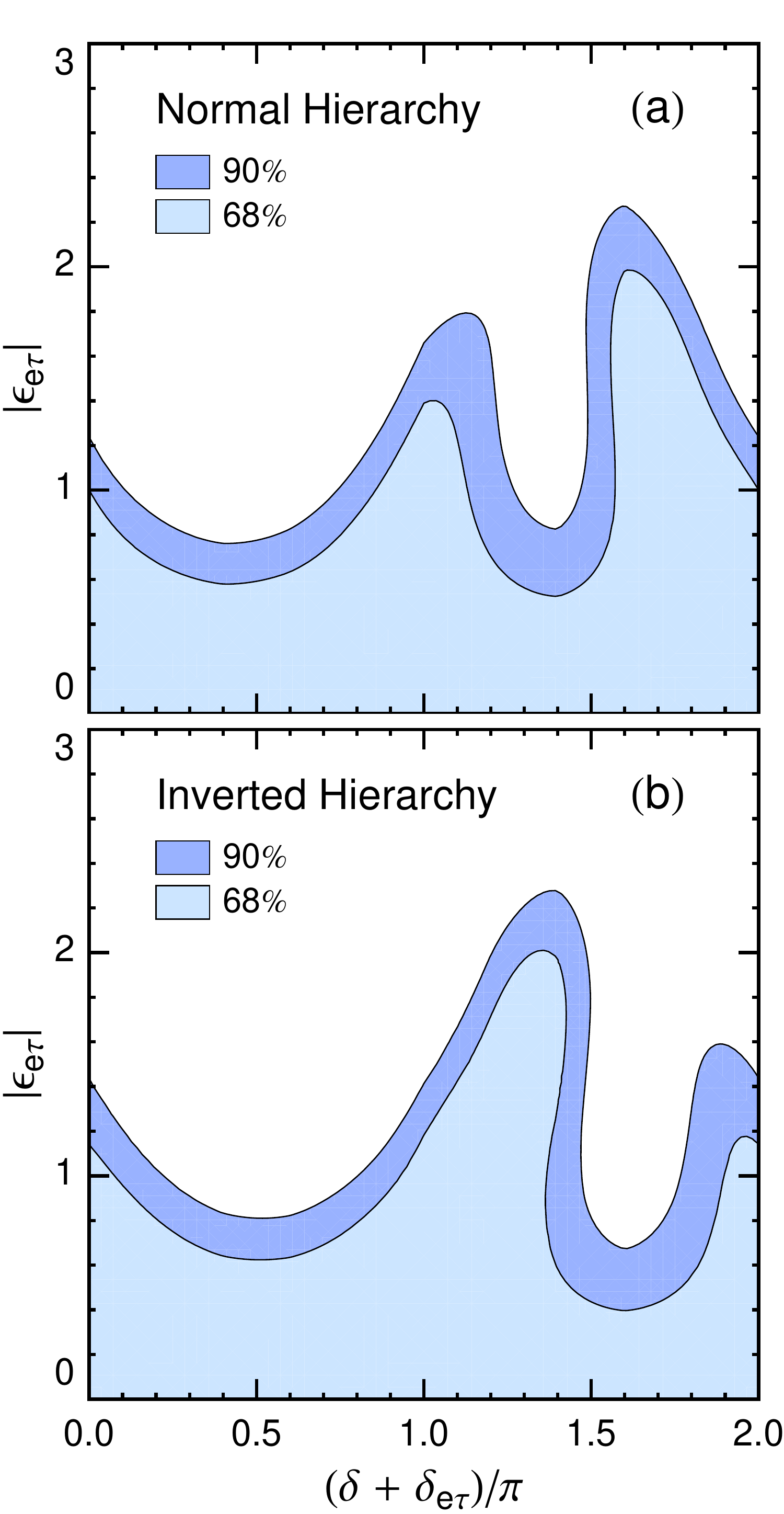}
\caption{\small Allowed-region contours of $\Delta \chi^2$ from fitting to three-flavor neutrino oscillations with 
$|\epsilon_{e \tau}| e^{i\delta_{e \tau}}V_{e}$ as the only operative NSI matter potential.   The regions allowed to
$|\epsilon_{e\tau}|$ and ($\delta + \delta_{e \tau}$)
are shown separately for the NH (Fig. 2a) and IH (Fig. 2b) neutrino mass hierarchies.   In each Figure,
values within the shaded region bounded by the upper (lower) border are allowed by the fit at $90\%$ ($68\%$) C.L. 
}
\label{fig:Delta-Epsilon-NH-IH}
\end{figure}

\subsection{$|\epsilon_{e \tau}| e^{i\delta_{e \tau}}$ as sole operative NSI}

For our second scenario we continue to treat $\epsilon_{e\tau}$ as the sole operative NSI, however we now treat it as a 
complex amplitude by including both its magnitude $|\epsilon_{e\tau}|$ and its CP phase $\delta_{e \tau}$
in the fit.   Additonally we allow the Dirac  CP phase to be operative.   
At T2K and MINOS baselines the contribution from
solar scale oscillations can be regarded as a perturbation, and a simplification arises in the limit of no solar scale which can
be harnessed to good effect.    As has been known for some time,
the phases $\delta$ and $\delta_{e \tau}$, in the limit $\alpha \rightarrow 0$,  
only appear in the $\nue$ appearance probability as the sum $\delta + \delta_{e \tau}$~\cite{ref:Yasuda, ref:Ota-2002}.
The verity of this assertion can be discerned in part by considering the oscillation probability of Eq.~\eqref{eq:approx-P}.   In Eq.~\eqref{eq:approx-P},
the first term contains ($\delta + \delta_{e \tau}$) as the phase of the amplitude, $f$, and the third term contains the same phase combination within an oscillatory cosine; 
otherwise the second and fourth terms vanish in the limit $\alpha \rightarrow 0 $, and the fifth and sixth terms are devoid of phases.

 With careful consideration of the $T_{3}$ amplitude of Eq.~\eqref{eq:Define-T3} and of the terms of $\mathcal{P}(\nu_\mu \rightarrow \nu_e)$ in which it
 enters ($T_{3}T_{1,2}^{*}+T_{3}^{*}T_{1,2}$ and $|T_{3}|^{2}$), the fact that the CP phases only appear in the $\alpha \rightarrow 0$ limit 
 as the sum $(\delta + \delta_{e \tau})$ can be
seen to hold exactly for $\nue$ appearance in constant-density matter.

Since the null solar-scale limit identifies ($\delta + \delta_{e \tau}$) to be the predominant source of phase in $\mathcal{P}(\nu_\mu \rightarrow \nu_e)$,
we express all phases within $\mathcal{P}(\nu_\mu \rightarrow \nu_e)$ in terms of the sum and the difference $\delta \pm \delta_{e \tau}$.    We then use
the sum-of-phases together with $|\epsilon_{e\tau}|$ as fit parameters, and marginalize over the difference-of-phases.  
As was done for the fit of Fig.~\ref{fig:Delta-chisq-vs-eps-etau}, we also marginalize over the $\theta_{23}$ and
$\theta_{13}$ mixing angles.    The $\chi^2$ fit identifies the regions allowed to the values $|\epsilon_{e\tau}|$ and ($\delta + \delta_{e \tau}$) as shown
in Fig.~\ref{fig:Delta-Epsilon-NH-IH}.    Figure~\ref{fig:Delta-Epsilon-NH-IH}a
shows the result of fitting to the NH;  the result for IH is shown in Fig.~\ref{fig:Delta-Epsilon-NH-IH}b.   Within each plot, the parameter regions
allowed by the fit at 68$\%$ and $90\%$ C.L.  are the shaded areas bounded by the lower and upper borders respectively.

For either hierarchy, there are sizable intervals for the sum-of-phases wherein $|\epsilon_{e\tau}|$ 
is constrained at $90\%$ C.L. to values distinctly smaller than
the limit obtained with our fit result of Eq.~\eqref{Real-e-tau-scenario}.    The improved constraints for $|\epsilon_{e\tau}|$ in Fig.~\ref{fig:Delta-Epsilon-NH-IH}
are made possible by allowing ($\delta + \delta_{e \tau}$) to be a fit parameter; the marginalization of the phase $\delta$ for the fit of Fig.~\ref{fig:Delta-chisq-vs-eps-etau}
effectively selects phases from regions of large excursion in $|\epsilon_{e\tau}|$ as appear in Figs.~\ref{fig:Delta-Epsilon-NH-IH}a,b.

\begin{figure}[!htb]
\includegraphics[width=8.5cm]{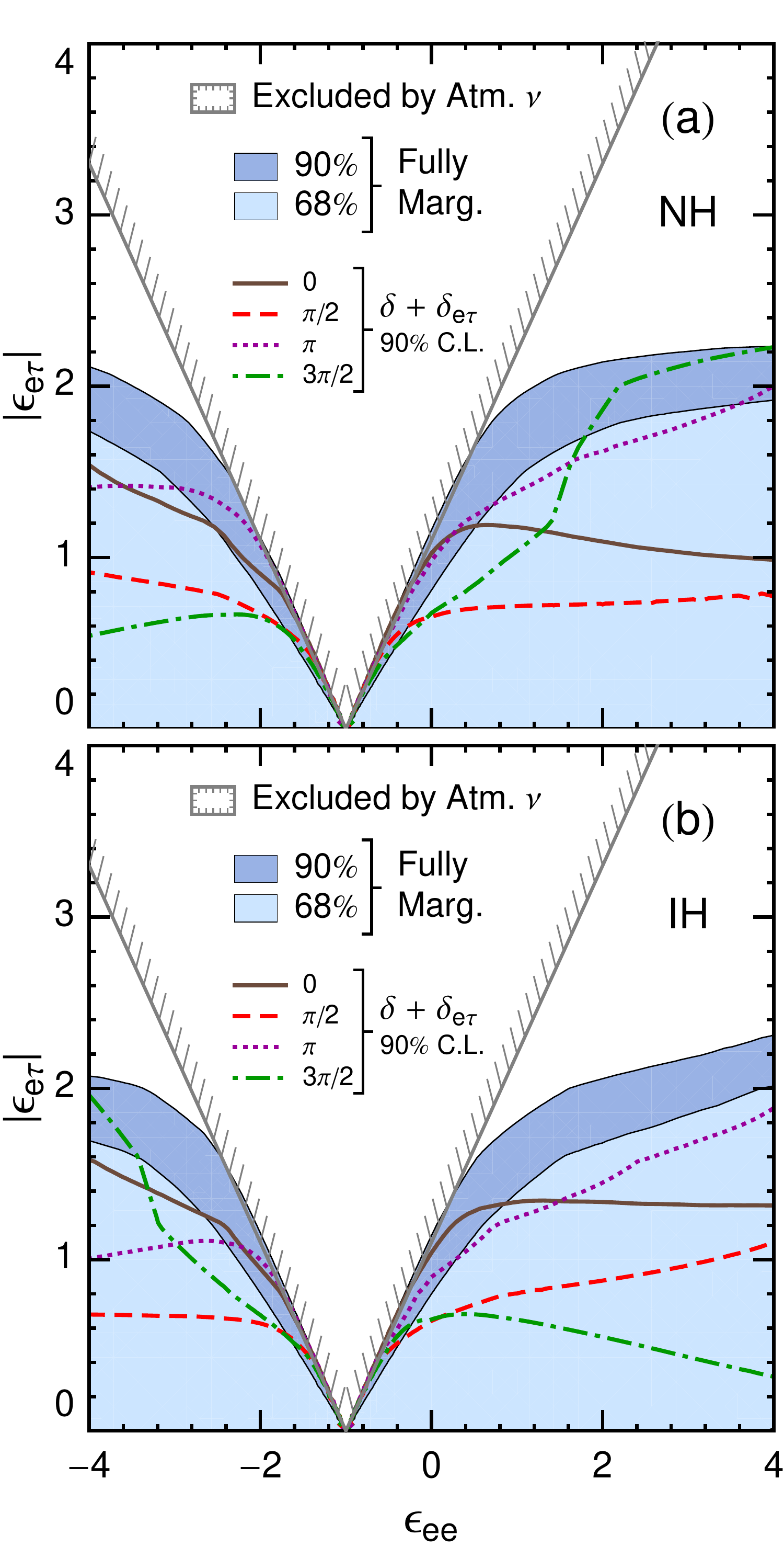}

\caption{\small  Allowed and excluded regions in the plane of $|\epsilon_{e \tau}|$ versus $\epsilon_{ee}$ for (a) NH and for (b) IH.    Shaded areas denote parameter regions allowed upon fitting to T2K and MINOS data, with marginalization over 
$\delta$ and $\delta_{e \tau}$, as well as over $\theta_{23}$ and $\theta_{13}$.   The V-shaped interior region (see Eq.~\eqref{eq:Atm-approx-bound}) is excluded on the basis of atmospheric neutrino and K2K data~\cite{ref:Yasuda}.    The fit sensitivity to CP phases is indicated using separate fits 
in which the value of $\delta + \delta_{e \tau}$ is fixed to, e.g. $0, \frac{\pi}{2}, \pi,$ or $\frac{3\pi}{2}$.  These give rise to upper boundaries at 90$\%$ C.L. for $|\epsilon_{e\tau}|$ as shown by the four curves in each plot
(solid, dashed, dotted, and dot-dashed respectively).}
\label{fig:NH-et-ee}
\end{figure}

\subsection{$\epsilon_{e \tau}$, $\epsilon_{ee}$, $\epsilon_{\tau\tau}$ with atmospheric constraints}

Given that current limits for $\epsilon_{ee}$ and $\epsilon_{\tau\tau}$ are even less stringent than those
for $\epsilon_{e \tau}$, full coverage of the possibilities requires that all three of these NSI be treated as operative.
Then $|\epsilon_{e \tau}|$, $\epsilon_{ee}$, $\epsilon_{\tau\tau}$, and ($\delta + \delta_{e \tau}$) will have
significant roles in the fit; on the other hand our data only consists of  three ``bins" of signal rates.   We
are thus motivated to utilize two observations gleaned from analysis of this same NSI scenario using the atmospheric neutrino data.

The first observation is that the allowed region of NSI couplings is well-characterized by an 
analytic expression~\cite{ref:FLM04, ref:FL05}:
\begin{equation}\label{eq:Atm-NSI-contour}
   \epsilon_{\tau\tau} \simeq  |\epsilon_{e \tau}|^{2}/(1 +  \epsilon_{ee}) .
\end{equation}
Relation \eqref{eq:Atm-NSI-contour} is implied by the requirement that oscillations with our three NSI couplings be consistent with
the high-energy atmospheric neutrino data~\cite{ref:Yasuda}.    For our final fit we assume relation $\eqref{eq:Atm-NSI-contour}$ to express
an equality;  with this assumption $\epsilon_{\tau\tau}$ can be expressed in terms of $|\epsilon_{e \tau}|$ and $\epsilon_{ee}$, thereby
reducing the number of NSI fit parameters.     Of course, we could as well use Eq.~$\eqref{eq:Atm-NSI-contour}$ to eliminate
$\epsilon_{ee}$ instead of $\epsilon_{\tau\tau}$; both approaches have been pursued in the literature~\cite{ref:Yasuda, ref:BOS}.

The second observation is an approximate bound deduced from atmospheric neutrino data~\cite{ref:Yasuda, ref:FL05}:
\begin{equation}\label{eq:Atm-approx-bound}
   |\epsilon_{e \tau}| \leq 1.1 \times |1 + \epsilon_{ee}| .
\end{equation}
In the following we use relation \eqref{eq:Atm-approx-bound} in conjunction with exclusion curves obtained from fitting to establish constraints
for $|\epsilon_{e \tau}|$.    For this purpose relation \eqref{eq:Atm-approx-bound} is very useful, for it eliminates a narrow region
of otherwise viable solutions for which $(1 + \epsilon_{ee}) \sim 0$ and hence $\epsilon_{\tau\tau}$, via Eq.~\eqref{eq:Atm-NSI-contour}, 
can be exceedingly large. 

For our final set of $\chi^{2}$ fits, we use  
$|\epsilon_{e \tau}|$ and $\epsilon_{e e}$ as fit parameters.   Figure~\ref{fig:NH-et-ee} displays our fit results as allowed regions 
in the plane of $|\epsilon_{e \tau}|$ versus $\epsilon_{e e}$ with the distinction made between the neutrino mass hierarchies, 
NH in Fig.~\ref{fig:NH-et-ee}a versus IH in Fig.~\ref{fig:NH-et-ee}b.  The straight-line
borders of the wedge-shaped region excluded by the atmospheric $\nu$ constraint as encoded by Eq.~\eqref{eq:Atm-approx-bound},
are superposed on each plot of Fig.~\ref{fig:NH-et-ee}.

Four separate fits have been carried out in which the sum of phases $(\delta + \delta_{e \tau})$ is fixed to a specific value in each fit, namely 
$0, \frac{\pi}{2}, \pi,$ and $\frac{3\pi}{2}$, while the difference in CP phases $(\delta - \delta_{e \tau})$ 
is marginalized over.  The outcomes are summarized by the four curves (solid-line, dashed, dotted, and dot-dash curves
respectively) which appear within the shaded areas of each plot.   The curves represent the boundaries which separate the regions  of allowed 
($|\epsilon_{e \tau}|$, $\epsilon_{e e}$) values (areas below the curves) from those which are excluded at 90\% C.L.   These results clearly 
suggest that limiting or measuring $|\epsilon_{e \tau}|$ at strengths below the MSW matter effect is a goal for the longer term.  At the baselines
considered here, even hierarchy discrimination in conjunction with a precision $\delta$ measurement does not assure that very restrictive
limits are achievable.

Our most realistic fits, however, are the ones for which both of the CP phases are included in the marginalization.    The outcomes of these latter
fits define the parameter regions allowed to $|\epsilon_{e \tau}|$ and $\epsilon_{e e}$ at 68$\%$ and 90$\%$ C.L. as depicted by the shaded
areas in Figs.~\ref{fig:NH-et-ee}a,b.     Thus the net effect of the recent T2K and MINOS data is to exclude those regions of relatively high  
$|\epsilon_{e \tau}|$which lie above the shaded allowed regions and are exterior to the region previously disfavored by the atmospheric neutrino data.

\subsection{Sensitivity to values of neglected NSI couplings}

It is appropriate at this stage to quantify the level of sensitivity that our $\epsilon_{e \tau}$ bounds may have, to perturbations
originating with neglected couplings operative within their allowed ranges.
For this purpose additional fits have been conducted for the NSI scenarios of Sections V.B and V.C, but with inclusion of one of 
$\epsilon_{\mu \mu}$, $\epsilon_{\mu \tau}$, or $\epsilon_{e \mu}$ in the fitting.   In these trials,  the
magnitudes of added NSI couplings were allowed to vary within the limits given in Sec. I.   With the $\epsilon_{\mu \tau}$ and $\epsilon_{e \mu}$ NSI, 
the CP phase degree-of-freedom was allowed for in the fitting.

From our ensemble of trial fits, we observe the $\epsilon_{e \tau}$ bounds reported in Sections V.B and V.C  to exhibit 
negligible sensitivity to $\epsilon_{\mu \mu}$.    With inclusion of $\epsilon_{\mu \tau}$, the bounds depicted in Fig.~\ref{fig:Delta-Epsilon-NH-IH} show a small sensitivity, 
with an upward shift of 4\% to the 90$\%$ C.L. boundary.     These outcomes are sensible as 
our analysis is based upon $\numu \rightarrow \nue$ appearance oscillations.    From the perspective of $\epsilon$-perturbation theory~\cite{ref:Minakata},
$\epsilon_{e \tau}$ occurs with strength $\epsilon^{2}$ in the transition probability whereas $\epsilon_{\mu \mu}$ and $\epsilon_{\mu \tau}$ occur with
strength $\epsilon^{3}$ and are therefore relatively suppressed.    

Our $\epsilon_{e \tau}$ bounds are somewhat more sensitive to $\epsilon_{e \mu}$ which is present in the oscillation probability with strength $\epsilon^{2}$.
For this NSI we use the 90$\%$ C.L. bound $|\epsilon_{e \mu}|< 0.33$ as reported by Ref.~\cite{ref:Biggio} which, in light of 
arguments by Ref.~\cite{ref:Yasuda} suggesting a smaller value, may be conservative.   Inclusion of $\epsilon_{e \mu}$ produces
modest overall elevation of contour boundaries for the $|\epsilon_{e \tau}|$ allowed regions;  the elevation
represents a relaxation of data constraints for $|\epsilon_{e \tau}|$ to the amount $\leq$ 6.5\% 
across the the upper borders of the shaded contours in Figs.~\ref{fig:Delta-Epsilon-NH-IH} and~\ref{fig:NH-et-ee}.  
Roughly characterized,  the boundary relaxation is the sum of two effects; a shift of $\sim 2.0\%$ arises from 
the range allowed to the modulus of $\epsilon_{e \mu}$, and a shift of $\sim4.5\%$ arises from variation of its CP phase.

\vskip 2pt
\section{Discussion}
Previous investigations of NSI matter effects in neutrino oscillations were hindered by lack of a measured value for the $\theta_{13}$ mixing angle.
The present work has availed itself of the recent delineation of $\theta_{13}$; it is the first study to use $\nue$ and $\anue$ appearance measurements from
accelerator long-baseline experiments, in conjunction with atmospheric-neutrino measurements, to obtain constraints for the complex $\epsilon_{e \tau}$
NSI coupling.  The constraints are expressed by the allowed regions in Figures~\ref{fig:Delta-Epsilon-NH-IH} and~\ref{fig:NH-et-ee}.
The limiting values for $|\epsilon_{e \tau}|$ vary according to the sum of the CP-violating phases $\delta + \delta_{e \tau}$ and 
according to the choice of neutrino mass hierarchy.    At $90\%$ C.L. the maximum value allowed to $|\epsilon_{e \tau}|$ varies from 0.7 
to 2.3, as shown by the minima and maxima in the allowed regions versus $\delta + \delta_{e \tau}$ for each mass hierarchy.    These values represent
an improvement upon the limit $|\epsilon_{e \tau}| < 3.0$ previously inferred from world data~\cite{ref:Biggio}.  
Nevertheless, the allowed range for $|\epsilon_{e \tau}|$ permitted by our analysis is still
relatively large, with the strength of the NSI potential exceeding that of the conventional MSW matter effect remaining as a viable possibility.   Our fit results of 
Fig.~\ref{fig:NH-et-ee} show that, with the flavor-diagonal NSI $\epsilon_{\tau \tau}$ expressed in terms of $|\epsilon_{e \tau}|$ and $\epsilon_{e e}$ using 
the atmospheric relation \eqref{eq:Atm-NSI-contour}, the $\nue$ appearance data does not provide any upper limits to the magnitude of the $\epsilon_{e e}$ NSI.
A $\nue$ appearance measurement to constrain $\epsilon_{e e}$ will likely require significantly more data, obtained
with detectors located at two different baselines such as in the 295 km and 1050 km baselines of the T2KK proposal~\cite{ref:Yasuda}.

In the near term, the T2K experiment is pursuing a precision measurement of $\mathcal{P}(\numu \rightarrow \nue)$ with $\numu$ exposures continuing
at higher beam power.    A $\nue$ appearance sample with sevenfold more events is projected for 2015~\cite{ref:Nakaya-T2K}.
For the MINOS experiment on the other hand, running with the NuMI neutrino beam in its low-energy configuration has completed.   A new round of
data taking with the MINOS detectors will commence in 2013 with the NuMI beam operating in medium-energy mode, for the MINOS+ experiment.
With medium-energy running, neutral-current
interactions yielding shower-like final states will occur at higher rates than was the case in MINOS exposures and so $\nue$ appearance
measurements in MINOS+ may not be feasible.    Medium-energy running of the NuMI beam however is optimal for $\nue$ appearance
measurements using NO$\nu$A.    The 810-km long baseline of NO$\nu$A is off-axis with respect to the NuMI beam; it receives a narrow-band neutrino flux for which
backgrounds originating from high energy neutral-current reactions are mostly suppressed.
It is the NO$\nu$A experiment whose observations of $\nue/\anue$ appearance rates and spectra, when taken together with new T2K measurements, hold promise 
for significant gains in delimiting NSI matter effects for neutrinos in propagation~\cite{ref:Friedland-pp2012}.   The discovery reach for NSI of the projected
T2K and NO$\nu$A exposures, when combined with measurements from the reactor experiments, has been examined in Ref.~\cite{ref:Kopp-2008}.
A point often made is that a degree of redundancy among the international suite of neutrino long-baseline accelerator and reactor experiments is useful for delineating the 
standard three-flavor oscillation framework.    With NSI matter effects for neutrinos in propagation included for consideration, multiple
measurements conducted at different baselines will be essential to affirming or ruling out the $\epsilon_{e \tau}$ non-standard interaction.

\vspace{+5pt}
\section*{Acknowledgments} \vspace{-8pt}
This work was supported by the United States Department of Energy under grant DE-FG02-92ER40702.

\section*{Appendix: Derivation of $\mathcal{A}(\nu_\mu \rightarrow \nu_e)$ including $\epsilon_{e \tau}$, $\epsilon_{e e}$, and $\epsilon_{\tau \tau}$ NSI}

Our derivation of an accurate $\mathcal{A}(\nu_\mu \rightarrow \nu_e)$ for the matter Hamiltonian of Eq.~\eqref{eq:matter-interactions-2}  for neutrinos propagating 
through a constant-density medium, proceeds as described in Sec. IV of Ref.~\cite{ref:Exact-Form}.   In brief, the conventional three-flavor Hamiltonian in
flavor basis $\hat H^{(\varphi)}$ is transformed to the propagation basis $\hat H^{(p)}$.    Upon re-phasing of the diagonal elements of $\hat H^{(p)}$ (with minor differences from the description in Ref.~\cite{ref:Exact-Form}), one arrives at the Hamiltonian of Eq.~\eqref{eq:elements-Hp}.
We separate $\hat H^{(p)}$ into an ``unperturbed" part, $\hat H_0^{(p)}$, plus an interaction potential, $\mathbb{\hat V}$:
\begin{equation} \label{eq:full_hamiltonian_prop_base}
   \hat H_0^{(p)} +  \mathbb{\hat V} = 
  \left(\begin{array}{ccc}
    -Q &  0 &  f \\
     0 & -G &  0 \\
     f^* &  0 & +Q
  \end{array}\right) + 
  \left(\begin{array}{ccc}
     0 &  r &  0 \\
     r^* &  0 &  b \\
     0 &  b^* & 0
  \end{array}\right).
\end{equation}
We then define an Interaction Picture:
\begin{equation} \label{eq:interaction}
  \vec \nu^{(I)}(t) = e^{i \hat H_0^{(p)}t} \vec \nu^{(p)}(t),\,\,
  \vec \nu^{(p)}(t) = e^{-i \hat H_0^{(p)}t} \vec \nu^{(I)}(t)  ~,
\end{equation}
so that
\begin{equation} \label{eq:state_evolution_inter_2}
  i \frac{d}{dt} \vec \nu^{(I)}(t) 
    =  \hat V_I \cdot \vec \nu^{(I)} (t)
\end{equation}
where
\begin{equation} \label{eq:V_I}
  \hat V_I(t) = e^{i \hat H_0^{(p)}t} \cdot \mathbb{\hat V} \cdot e^{-i \hat H_0^{(p)}t}.
\end{equation}

Our approach is to solve for the time evolution operator 
in the Interaction Picture $\hat U_I(t=\ell,\,t = 0)$:
\begin{equation} \label{eq:interaction_time}
  \vec \nu^{(I)}(t) = \hat U_I(t,0)  \cdot \vec \nu^{(I)}(0).
\end{equation}
Substitution of Eq.~(\ref{eq:interaction_time}) into Eq.~(\ref{eq:state_evolution_inter_2}) 
yields the wave equation which governs $\hat U_I (t, 0)$ :
\begin{equation} \label{eq:wave_eq_time_evolve_op}
  i \frac{d}{dt} \hat U_I (t, 0) = \hat V_I (t) \hat U_I (t, 0).
\end{equation}

To obtain $\hat V_I(t)$ we require the matrix representation (in propagation basis) of the unitary
operator forms $\mathrm{exp}(\pm i \hat H_0^{(p)} t)$. 
From $\hat H_0^{(p)}$ we extract the reduced matrix
\begin{equation}
\begin{split}
  \hat H_{0, R}^{(p)}
    &= \left(
      \begin{array}{cc}
        -Q & f_0 + i f_1 \\
        f_0 - i f_1 & +Q
      \end{array} \right) ,
\end{split}
\end{equation}
where $f_{0}$ and $f_{1}$ designate the real and imaginary parts of the element $(\hat H_{0}^{(p)})_{13} \equiv f$.
The reduced matrix can be decomposed using Pauli matrices,
\begin{equation}
\begin{split}
  \hat H_{0, R}^{(p)}
    &= f_0 \hat \sigma_x - f_1 \hat\sigma_y - Q \hat \sigma_z \\
    &= \vec N \cdot \vec \sigma, \;\;
    \mathrm{where}\;\; \vec N = (f_0, -f_1, -Q).
\end{split}
\end{equation}
We have $|\vec N| = N = \sqrt{f_0^2 + f_1^2 + Q^2}$.  Its unit vector $\hat n$ defines the axis of rotation
in the reduced (spinor) space,
\begin{equation}\label{eq:define-vector-n}
  \hat n = (n_x,\; n_y,\; n_z)
         = \frac{1}{\left(|f|^2+Q^2\right)^{\frac{1}{2}}} \left(f_0, -f_1, -Q\right).
\end{equation}

Designating the angle of rotation with 
\begin{equation}\label{eq:Define-phi}
\phi \equiv N \ell, 
\end{equation}
we use the spinor identity 
\begin{equation}
\begin{split}
  e^{i \vec\sigma \cdot \hat n \phi}
    &= \left(
      \begin{array}{cc}
        \cos\phi + i n_z \sin\phi &
        (i n_x + n_y) \cdot \sin\phi \\
        (i n_x - n_y) \cdot \sin\phi &
        \cos\phi - i n_z \sin\phi        
      \end{array} \right)
\end{split}
\end{equation}
and furthermore define
\begin{equation}\label{eq:define-gamma-beta}
\begin{split}
  \gamma &\equiv \cos\phi + i n_z \sin\phi, \\
  \beta &\equiv  \beta_x + i \beta_y ,~ \beta_x  \equiv  n_x \sin\phi , ~ \beta_y  \equiv  n_y \sin\phi .
\end{split}
\end{equation}
Then $ i\beta = i\beta_x - \beta_y$ and $ i\beta^* = i\beta_x + \beta_y $, and we have 
\begin{equation}
  e^{i \hat H_{0,R}^{(p)}\ell}
    = e^{i \vec\sigma \cdot \hat n(N\ell)}
    = \left(
      \begin{array}{cc}
        \gamma & i\beta^* \\ i\beta & \gamma^*
      \end{array} \right).
\end{equation}
Thus in the propagation basis we may write
\begin{equation}
  e^{i \hat H_0^{(p)}\ell}
    = \left(
      \begin{array}{ccc}
        \gamma & 0 & i\beta^* \\
        0 & e^{-iG\ell} & 0 \\
        i \beta & 0 & \gamma^*
      \end{array} \right).
\end{equation}

To move the formalism to the Interaction Picture, we evaluate
\begin{equation}
\begin{split}
 \hat V_I (\ell)
    &= e^{i \hat H_0^{(p)}\ell} \cdot \hat \V \cdot e^{-i \hat H_0^{(p)}\ell} \\
    &=  \left(
      \begin{array}{ccc}
        0 & u & 0 \\
        u^* & 0 & v \\
        0 & v^* & 0
      \end{array} \right) ,
 \end{split}
\end{equation}
where the complex elements of $\hat V_I (\ell)$ are
\begin{equation}\label{eq:define-u-and-v}
\begin{split}
  u \equiv (\gamma r + i \beta^* B) e^{iG\ell},\;\;
  v \equiv (\gamma B - i \beta^* r^*) e^{-iG\ell}.
\end{split}
\end{equation}
Now
\begin{equation}
  \left( \hat V_I(\ell) \right)^2
    = \left(
      \begin{array}{ccc}
        |u|^2 & 0 & uv \\
        0 & |u|^2 + |v|^2 & 0 \\
        (uv)^* & 0 & |v|^2
      \end{array} \right).
\end{equation}
The real-valued expression $( |u|^2 + |v|^2)$ recurs upon taking higher
integer powers of $\hat V_I (\ell)$.   It is readily reduced to $(|r|^2 + |b|^2)$, previously designated as $\eta^2$ in Eq.~\eqref{define-eta-bar}:
\begin{equation}\label{eq:Define-eta-sqrd}
 \eta^2 =  |u|^2 + |v|^2 =  |r|^2 + |b|^2 .
\end{equation}

The exponentiation of $\hat V_I(\ell)$ into $e^{-i \hat V_I \ell}$ proceeds as in 
Ref.~\cite{ref:Exact-Form}.  We obtain
\begin{equation}\label{eq:Identity-VI}
\begin{split}
  e^{-i\hat V_I \ell}
    = \hat \I - \left(\frac{\hat V_I}{\eta}\right)^2(1-\cos(\eta\ell))
      - i \frac{\hat V_I}{\eta} \sin(\eta\ell).
\end{split}
\end{equation}

We define
\begin{equation}\label{eq:define-theta-baru-barv}
  \theta \equiv \eta\ell,\;\;
  \bar u \equiv \frac{u}{\eta}, \;\;
  \bar v \equiv \frac{v}{\eta},
\end{equation}
and write the diagonal elements of Eq.~\eqref{eq:Identity-VI} as
\begin{equation} \label{eq:def-1}
  D_u \equiv 1 - 2 |\bar u|^2 \cdot \sin^2 \frac{\theta}{2},\;
  d \equiv \cos\theta,\;
  D_v \equiv 1 - 2|\bar v|^2 \cdot \sin^2 \frac{\theta}{2}.
\end{equation}
For the off-diagonal elements we define
\begin{equation} \label{eq:def-2}
  w \equiv \bar u \sin\theta,\;\;
  p \equiv -2 \bar u \bar v \sin^2 \frac{\theta}{2},\;\;
  k \equiv \bar v \sin\theta ~.
\end{equation}
Then the evolution operator in the Interaction Picture is
\begin{equation}
\begin{split}
 \hat U_I(\ell,0)
    = e^{-i \hat V_I \ell}
    = \left( \begin{array}{ccc}
        D_u & -iw & p \\
        -i w^* & d & -ik \\
        p^* & -ik^* & D_v
      \end{array} \right)
\end{split}
\end{equation}
and, in the propagation basis, it becomes
\begin{equation}
\begin{split}
 &\hat U^{(p)}(\ell,0)
    = e^{-i \hat H_0^{(p)} \ell} \cdot \hat U_I(\ell, 0) \\
    &= \left( \begin{array}{ccc}
        \gamma^* D_u - i\beta^* p^* & \gamma^*(-iw)-\beta^*k^* & \gamma^* p - i \beta^* D_v \\
        -iw^* e^{iG\ell} & d e^{iG\ell} & (-ik)e^{iG\ell} \\
        \gamma p^* - i \beta D_u & \gamma(-ik^*)-\beta w & \gamma D_v - i\beta p
      \end{array} \right).
\end{split}
\end{equation}

Finally, returning to flavor basis
\begin{equation}
\begin{split} \label{eq:U_flavor_basis}
 \hat U^{(\varphi)}(\ell,0)
    &= ( \hat R_1 \hat \I_\delta) \cdot \hat U^{(p)}(\ell, 0) \cdot (\hat \I_{-\delta} \hat R_1^T)
\end{split}
\end{equation}
we obtain
\begin{equation}\label{eq:Element-12}
\begin{split}
& (\hat U^{(\varphi)})_{12} = \Amplitude = \\
&~c_{23} U_{12}^{(p)} + s_{23} U_{13}^{(p)} e^{-i\delta} = \\
&~(-i) c_{23} (\gamma^* w - i \beta^* k^*) + s_{23} \gamma^* p e^{-i\delta}
               - i s_{23} \beta^* D_v e^{-i\delta}.
\end{split}
\end{equation}
We insert expressions \eqref{eq:def-1} and \eqref{eq:def-2} for the elements of $k$, $p$, and $D_{v}$ into the last line of Eq.~\eqref{eq:Element-12}
and re-arrange the order of the terms to obtain
\begin{equation}\label{eq:A_3_terms-summary}
\begin{split}
  \Amplitude
    &= (-i) s_{23} \beta^* e^{-i\delta} \\
    &\;\;\; + (-i) c_{23} \left[ \gamma^* \bar u - i \beta^* \bar v^* \right] \sin\theta \\
    &\;\;\; + 2 s_{23} \left[ i \beta^* |\bar v|^2 - \gamma^* \bar u \bar v \right]
              \cdot \sin^2 \frac{\theta}{2} \cdot e^{-i\delta}. 
\end{split}
\end{equation}

The three terms of Eq.~\eqref{eq:A_3_terms-summary} correspond, respectively, to the terms $T_1 + T_2 + T_3$ of Eq.~\eqref{eq:Amplitude-Sum}. 

\smallskip

For the first term, $T_1$, we use Eqs.~\eqref{eq:define-gamma-beta}, \eqref{eq:Define-phi}, and \eqref{eq:define-vector-n} to write
\begin{equation}
  \beta^*  = (n_x - i n_y)\sin(N\ell) = \frac{1}{N}(f_0 + if_1)\sin(N\ell) ;
\end{equation}
the term reduces immediately to 
\begin{equation} \label{eq:T1-derive}
  T_1 = (-i) s_{23} \frac{f}{N} \sin(\bar N \Delta) \cdot e^{-i\delta} .
\end{equation}

Considering the second term $T_2$, we use Eqs.~\eqref{eq:define-u-and-v} and \eqref{eq:define-theta-baru-barv} to insert
\[
  \bar u = \frac{1}{\eta} (\gamma r + i \beta^* B)e^{iG\ell},\;\;\;
  \bar v^* = \frac{1}{\eta} (\gamma^* B + i \beta r) e^{iG\ell},
\]
and find that it reduces to
\begin{equation}\label{eq:T2-derive}
  T_2 = (-i) c_{23} \frac{r}{\eta} \cdot \sin(\bar \eta \Delta) \cdot e^{i\bar G \Delta}.
\end{equation}

\smallskip
The remaining term is
\begin{equation}
  T_3 = 2 s_{23} \left[ i\beta^* |\bar v|^2 - \gamma^* \bar u \bar v \right]
        \cdot \sin^2 \frac{\theta}{2} \cdot e^{-i\delta}.
\end{equation}
The expression within the bracket reduces to
 $\frac{1}{\eta^2} \left[ -\gamma rb + i \beta^* |r|^2 \right] $
so that
\begin{equation}
  T_3 = -2 s_{23} \left\{ \left( \frac{rb}{\eta^2}\right) \cdot \gamma
                          -i \left( \frac{|r|^2}{\eta^2} \right) \cdot \beta^* \right\}.
\end{equation}
Substitution of $\gamma$ and $\beta^*$ from Eq.~\eqref{eq:define-gamma-beta} yields
\begin{equation} \label{eq:T3-derive}
\begin{split}
  T_3 &  = (-2 s_{23}) \cdot \sin^2 \frac{\theta}{2} \cdot e^{-i\delta} \cdot     \\
       & \left\{ \frac{rb}{\eta^2} \cos\phi + \left[ \frac{-|r|^2 n_y}{\eta^2}
          + i \left( \frac{rb}{\eta^2} \cdot n_z - \frac{|r|^2}{\eta^2} n_x \right) \right]
        \sin\phi \right\}.
\end{split}
\end{equation}
Within the curly brackets on the right-hand side, the factors multiplying $\cos \phi$ comprise
the complex function $S_{1}$ of Eq.~\eqref{eq:S1-def}, and the expression which multiplies $\sin \theta$ is the complex function
$S_{2}$ of Eq.~\eqref{eq:S2-def}.  Thus Eq.~\eqref{eq:T3-derive} coincides with Eq.~\eqref{eq:Define-T3} for $T_{3}$. 

\smallskip

With Eqs.~\eqref{eq:T1-derive}, \eqref{eq:T2-derive}, and \eqref{eq:T3-derive} we have shown that $T_{1}$, $T_{2}$, and $T_{3}$ have the forms as previously
specified in Eqs.~\eqref{eq:Define-T1}, \eqref{eq:Define-T2}, and \eqref{eq:Define-T3} of Sec.~\ref{sec:Specify-Ts}.   
The transition amplitude $\Amplitude$ of Eq.~\eqref{eq:Amplitude-Sum} is thus completely specified.

\end{document}